\documentstyle[12pt]{article}

\def\dalemb#1#2{{\vbox{\hrule height .#2pt
        \hbox{\vrule width.#2pt height#1pt \kern#1pt
                \vrule width.#2pt}
        \hrule height.#2pt}}}

\let\a=\alpha   \let\d=\delta \let\e=\epsilon
 \let\h=\eta \let\th=\theta  
 \let\m=\mu \let\n=\nu   
\let\s=\sigma \let\t=\tau    
 
      \let\G=\Gamma  \let\Th=\Theta

\def\nn{\nonumber} \def\bd{\begin{document}} \def\ed{\end{document}}
\def\ds{\documentstyle} \let\fr=\frac \let\bl=\bigl \let\br=\bigr
\let\Br=\Bigr \let\Bl=\Bigl
\let\bm=\bibitem
\let\na=\nabla
\def\tU{{\widetilde U}}
\let\pa=\partial \let\ov=\overline
\def\ie{{\it i.e.\ }}
\newcommand{\be}{\begin{equation}}
\newcommand{\ee}{\end{equation}}
\def\ba{\begin{array}}
\def\ea{\end{array}}
\def\ft#1#2{{\textstyle{{\scriptstyle #1}\over {\scriptstyle #2}}}}
\def\fft#1#2{{#1 \over #2}}
\def\F#1#2{{ F_{#1}^{(#2)} }}
\def\cF#1#2{{ {\cal F}_{#1}^{(#2)} }}

\def\R{{\bf R}}
\def\sst#1{{\scriptscriptstyle #1}}
\def\oneone{\rlap 1\mkern4mu{\rm l}}
\def\e7{E_{7(+7)}}
\def\td{\tilde}
\def\wtd{\widetilde}
\def\im{{\rm i}}
\def\bog{Bogomol'nyi\ }
\newcommand{\ho}[1]{$\, ^{#1}$}
\newcommand{\hoch}[1]{$\, ^{#1}$}
\newcommand{\bea}{\begin{eqnarray}}
\newcommand{\eea}{\end{eqnarray}}
\newcommand{\ra}{\rightarrow}
\newcommand{\lra}{\longrightarrow}
\newcommand{\Lra}{\Leftrightarrow}
\newcommand{\ap}{\alpha^\prime}
\newcommand{\bp}{\tilde \beta^\prime}
\newcommand{\cB}{{\cal B}}
\newcommand{\cO}{{\cal O}}
\newcommand{\vecx}{\vec{x}}
\newcommand{\vecy}{\vec{y}}
\newcommand{\vecp}{\vec{p}}
\newcommand{\vecq}{\vec{q}}
\newcommand{\tr}{{\rm tr} }
\newcommand{\Tr}{{\rm Tr} }
\newcommand{\NP}{Nucl. Phys. }

\def\sst#1{{\scriptscriptstyle #1}}
\def\0{{\sst{(0)}}}
\def\1{{\sst{(1)}}}
\def\2{{\sst{(2)}}}
\def\3{{\sst{(3)}}}
\def\4{{\sst{(4)}}}
\def\5{{\sst{(5)}}}
\def\6{{\sst{(6)}}}
\def\7{{\sst{(7)}}}
\def\8{{\sst{(8)}}}
\def\ve{\varepsilon}
\def\vf{\varphi}
\def\F{\Phi}
\def\e{\epsilon}
\def\wg{\wedge}

\newcommand{\tamphys}{\it Center for Theoretical Physics,\\
Texas A\&M University, \\College Station, Texas 77843}

\newcommand{\auth}{ M. Mihailescu
\footnote{New address is  \\
\indent                          CARB\\
\indent                   University of Maryland Biotechnology Institute\\ 
\indent                   Rockville, MD 20850}, I.Y. Park, and T.A. Tran}

\def\thb{\bar{\theta}}
\def\barp{\bar{p}}
\def\barq{\bar{q}}
\def\barc{\bar{c}}
\def\bard{\bar{d}}
\def\e{\epsilon}
\def\h{\eta}

\def\th{\theta}\def\Th{\Theta}\def\vth{\vartheta}

\def\umu{{\underline \mu}}
\def\unu{{\underline \nu}}

\begin{document}
\begin{flushright}
\hfill{CTP TAMU-33/00}\\
\hfill{hep-th/0011079}\\
\hfill{}\\
\end{flushright}

\vspace{20pt}

\begin{center}
{\large {\bf D-branes as Solitons of 
              an ${\cal N}=1, D=10$ Non-commutative Gauge Theory }}

\vspace{30pt}
\auth

{\tamphys}
     
\vspace{40pt}
         
\begin{abstract}
We consider a Dp brane within a D9 brane in the presence of a B-field 
whose polarization is {\em transverse} to the Dp brane. To be definite, 
we take a D3-D9 system. It is observed that the system has the same 
pattern of supersymmetry breaking as that of a soliton 
of the six dimensional non-commutative gauge theory that is
obtained by dimensional reduction of an ${\cal N}=1, D=10$ gauge
theory. These results indicate that the soliton solution 
is the low energy realization of a D3 brane in a D9 brane with a
transverse B-field, hence can be viewed as a generalization of the
previous results in the literature where similar observations were
made for lower codimensional cases. 

\end{abstract}

\end{center}

\newpage

\section{Introduction}
D-branes appear when one T-dualizes open strings with fully Neumann
boundary conditions.
Since the duality connects two equivalent 
descriptions, it is worth asking whether open string
theory has a different way of incorporating D-branes. 
The most natural candidate that could provide such incorporation may be 
the soliton solutions of the low energy effective action of the open
string theory. 

Progress toward different realizations of D-branes has been 
made\footnote{Earlier discussions of D-branes as instantons can be
found, for example, in \cite{w2,d}.} in the
frame work of open bosonic string field theory \cite{witten} and 
non-commutative gauge theory \cite{cds,dh}. In particular, it was
argued that D-branes can be constructed as solitons of open string 
field theory \cite{s1,hk,djmt}. A similar claim has also been made for the
soliton solutions of non-commutative
gauge theories \cite{nekrasov,gms,b,agms,hklm,hkl,gn}. In this note, we study
the pattern of supersymmetry breaking of such non-commutative solitons.

To be definite, we consider a D3-D9 system although various other
 systems can be analyzed similarly.  (Various properties of generic p-p'
 systems were studied in \cite{cimm} that has some related       
discussions.\footnote{We thank E. Witten for
 bringing this paper to our attention.} ) We consider oriented open
 superstring description of
D-branes \cite{lw,cz} using the GS formalism \cite{gs}.  
 A constant B-field is turned on with the only non-zero components
{\em transverse} to the D3 brane. The boundary conditions of the system
 impose a constraint on the parameters of the supersymmetry 
 transformations. The constraint equation in an 
appropriate limit is to be compared with the corresponding gauge
 theory result. 

The low energy limit of the open superstring theory is 
an ${\cal N}=1$ gauge theory in $D=10$. We dimensionally reduce the theory 
to $D=6$.  
With the same B-field configuration as that of the D3-D9 system, one gets 
six dimensional 
non-commutative gauge theory. Following \cite{agms}, soliton
solutions are written down. By a trivial lifting procedure, they can
also be considered as solutions of the ten dimensional gauge theory
that one started with. We examine the supersymmetry preserving conditions and 
observe that the supersymmetry breaking pattern is equivalent to the 
constraint equation obtained for the D3-D9 system. This can be viewed as a 
generalization of the previous results in the literature where similar
observations were made for lower codimensional cases. 
\\ 
 
The rest of the paper is organized as follows: In section 2, we
consider a D3-D9 brane system with the only non-zero components
transverse to the D3 brane. The boundary condition at one
end of the open string is different from the one at the other end. This
makes it necessary, for any possible residual supersymmetry, to impose
an extraconstraint among supersymmetry parameters in addition to the  
ordinary one that breaks one half of the supersymmetry. In section 3, we
consider a ${\cal N}=1$ supersymmetric
gauge theory in $D=10$ and dimensionally reduce it to six dimensions
that have the non-zero B-components. Therefore the six dimensional theory is 
non-commutative. We show that the supersymmetry breaking pattern for
the soliton solutions is equivalent to that of the D3-D9 system.
Although the supersymmetry gets completely broken for a generic configuration
of the B-field, one can choose the values of the B-components such that the system has residual supersymmetry. It is shown
in section 4 that there are no tachyons for those values of the B-field.
We end with conclusions in section 5.

\section{Open Superstring Analysis}

Consider an open superstring in a  
flat ten dimensional background.
We turn on constant B field and impose Dirichlet 
boundary conditions appropriate for a given D-brane configuration. We briefly 
review \cite{lw,cz} for a D3 brane case where the authors considered 
a $B$-field polarized along the brane direction. Then we move to the case of
transverse $B$-field.

In the Green-Schwarz formulation, the action with a B-field is given 
by\footnote{We denote the coordinate by  
$
M = (\m,m), \quad \m=0,...,3, \quad m=4,...,9 
$      }
\bea 
S &=& -\fr{1}{2\pi}\int d^2 \s \; \left(\;\sqrt{-g} g^{ij} 
       \Pi_i{}^M \Pi_j{}^N \eta_{MN} 
                  +2 i \e^{ij} \pa_i X^M 
                 ( \thb^1 \G_M \pa_j \th^1 
                        -\thb^2 \G_M \pa_j \th^2) \right. \nn\\
  & & \left.\hspace{1in}  - 2 \e^{ij} 
                       (\thb^1 \G^M\pa_i\th^1)(\thb^2 \G_M 
                \pa_j \th^2)+ \e^{ij} \pa_i X^M \pa_j X^N 
                               B_{MN} \;\right)
\label{action}
\eea
By taking the variations with respect to $X$ and $\th$, one can obtain
the following boundary terms,
\bea
&& \d X_M \left(\Pi_\s{}^M -i\thb^1\G^M\pa_\t\th^1
               +i\thb^2\G^M\pa_\t\th^2 \right) 
                 + \d X_M \pa_\t X^N B_N{}^M 
                   -i(\thb^A\G_M\d\th^A) \Pi_\s{}^M  \nn\\
&& + i\pa_\t X^M(\thb^1\G_M\d\th^1 -\thb^2\G_M\d \th^2)
    + ( \thb^1 \G^M \d \th^1 \thb^2 \G_M \pa_\t \th^2 
     -\thb^2 \G^M \d \th^2 \thb^1 \G_M \pa_\t \th^1)|_{\s=0,\pi}  
=0. \nn\\  \label{BC}
\eea
With the B-field whose non-zero components are only along the brane
directions, the equation above becomes
\bea
&&  \d X_M \left(\Pi_\s{}^M -i\thb^1\G^M\pa_\t\th^1
        +i\thb^2\G^M\pa_\t\th^2 \right) 
       + \d X_\mu \pa_\t X^\nu B_\nu{}^\mu 
       -i(\thb^A\G_M\d\th^A) \Pi_\s{}^M \nn \\
&& + i\pa_\t X^M(\thb^1\G_M\d\th^1 -\thb^2\G_M\d \th^2)
        + ( \thb^1 \G^M \d \th^1 \thb^2 \G_M \pa_\t \th^2 
          -\thb^2 \G^M \d \th^2 \thb^1 \G_M \pa_\t \th^1)|_{\s=0,\pi}  
          =0. 
\nn\\ \label{BC1}
\eea
The $X$-variation vanishes if one impose the following boundary conditions,
\bea 
&&  \pa_\t X^{m}= 0,  \quad
 \pa_\s X^\mu + \pa_\tau X^\nu B_\nu{}^\mu =0,
                \nn  \\
&&  \thb^1\G^\mu (\pa_\s +\pa_\t)\th^1 
      +\thb^2\G^\mu (\pa_\s -\pa_\t)\th^2 =0 \,,
\label{nbc3}  
\eea
The part containing $\d\th$ can be rewritten as 
\bea
& i \pa_\s X^{m} ( \thb^1 \G_{m} \d \th^1 +  \thb^2 \G_{m} \d
\th^2 ) 
+ i \pa_\t X^{\mu} [ (1+B)_{\mu\nu}\thb^2\G^\nu\d\th^2 
- (1-B)_{\mu\nu}\thb^1\G^\nu\d\th^1 ] \nn\\
&+\thb^1 \G_{M} \d \th^1  \thb^1 \G^{M} \pa_\t \th^1 
 - \thb^2 \G_{M} \d \th^2  \thb^2 \G^{M} \pa_\t \th^2=0. 
\label{thetaterms}
\eea
This can be satisfied by imposing the following relation between the
two $\th$'s,
\be\label{G}
\th^2 =P(B_{\parallel}) \; \th^1 \quad {\rm at} \ \s=0,\pi ,
\ee
where
\be \label{aga}
P(B_\parallel) \equiv
e^{(-{1\over2} Y^{\m\n} \G_{\m\n} \s_3)}
i \s_2 \,\G^{0 \cdots 3} 
\ee
with
\be
Y=\frac{1}{2}\ln \left(\frac{1-B}{1+B}\right). 
\ee

Now, we consider the D3-D9 system having a B-field along the transverse 
directions to D3. The boundary conditions are as follows,
\begin{center}
\begin{tabular}{ccccccccccc}
$\sigma$ & $X^\circ$ & $X^1$ & $X^2$ & $X^3$ & $X^4$ & $X^5$ & $X^6$ 
         & $X^7$ & $X^8$ & $X^9$  \\
$0$      & N &  N &  N &  N & D &D &D &D &D &D  \\
$\pi$    & N &  N &  N &  N & ND& ND& ND& ND& ND& ND  \\
\end{tabular}
\end{center}
where ND denotes the boundary condition of the type,
$\pa_\s X+\pa_\t X \cdot B $.
It is convenient to examine the boundary conditions at $\s=0$ and
$\s=\pi$ separately.\\

{\bf i) Boundary conditions at $\s=0$}\\
First consider boundary conditions at $\s=0$. We require the
following conditions:
\bea
&& \pa_\t X^{m} = 0,  \nn\\
&&    \pa_\s X^{\m}  = 0,  \nn\\
&& \thb^1\G^M (\pa_\s +\pa_\t)\th^1 
       +\thb^2\G^M (\pa_\s -\pa_\t)\th^2 =0 \,,
\label{sigma0}  
\eea
Two remarks are in order: first, it is important that the B-field does not
appear in the equations above whereas it does appear in the boundary
conditions at $\s=\pi$. This difference is
what induces an additional constraint on supersymmetry parameters as will be
discussed below.
Secondly, although we can replace  $\G^M$ in the third equation
by $\G^\mu$, we keep the full $\G$-matrices because at $\s=\pi$ we can
not make the same replacement. 
With these conditions, it is easy to show that $\d X$ part vanishes.
The remaining $\d \th$ part is as follows:
\bea 
& - i \pa_\s X^{m} ( \thb^1 \G_{m} \d \th^1 
              +  \thb^2 \G_{m} \d \th^2 ) 
           + i \pa_\t X^{\mu} [ \d_{\mu\nu}\thb^1\G^\nu\d\th^1 
                    - \d_{\mu\nu}\thb^2\G^\nu\d\th^2 ] \nn\\
&  +\thb^1 \G_{M} \d \th^1  \thb^1 \G^{M} \pa_\t \th^1 
     - \thb^2 \G_{M} \d \th^2  \thb^2 \G^{M} \pa_\t \th^2=0. 
\eea
which implies a relation between $\th^1$ and $\th^2$ is
\be
\th^2 =P_{\s=0}(B_\perp) \,\th^1 
\label{constraint1}
\ee
where
\be 
P_{\s=0}(B_\perp)=
 i \s_2 \,\G_{4 \cdots 9} 
\ee

{\bf i) Boundary conditions at $\s=\pi$}\\
The boundary conditions that remove
the $\d X$ part are,
\bea
&& \pa_\s X^{\m} = 0,  \nn\\
&& \pa_\s X^{m}+\pa_{\t} X^{n}B_{n}^{\;\;m}  = 0, \nn\\
&&              \thb^1\G^M (\pa_\s +\pa_\t)\th^1 
               +\thb^2\G^M (\pa_\s -\pa_\t)\th^2 =0 
\label{sigmapi}
\eea
Unlike (\ref{sigma0}), equation (\ref{sigmapi}) contains the B-field,
which will,
in turn, modify the constraint equation (\ref{constraint1}) accordingly.  
The remaining $\d \th$-part is 
\bea 
& i \pa_\s X^{\mu} ( \thb^1 \G_{\mu} \d \th^1 -  \thb^2 \G_{\mu} \d
\th^2 ) 
+ i \pa_\t X^{m} [ (-1+B)_{mn}\thb^2\G^{n}\d\th^2 
+ (1+B)_{mn}\thb^1\G^{n}\d\th^1 ] \nn\\
& +\thb^1 \G_{M} \d \th^1  \thb^1 \G^{M} \pa_\t \th^1 
 - \thb^2 \G_{M} \d \th^2  \thb^2 \G^{M} \pa_\t \th^2=0. 
\eea
The modified relation between the $\th$'s is
\be
\th^2 =P_{\s=\pi}(B_\perp) \th^1 
\ee
where
\be 
P_{\s=\pi}(B_\perp) =
e^{(-{1\over2} Y^{mn} \G_{mn} \s_3)}
i \s_2 \,\G^{4 \cdots 9} 
\ee
with
\be
Y=\frac{1}{2}\ln \left(\frac{1+B}{1-B}\right). 
\label{Y}
\ee
The action (\ref{action}) is invariant under the following
supersymmetry transformation
\bea
\d X^M &=& i\bar{\e}^A \G^{M}\th^A \nn\\
\d \th    &=& \e^A
\eea
In order for these transformations to be consistent with the boundary
conditions discussed above, it must be satisfied that
\be
\e^2=P_{\s=0}(B_\perp)\,\e^1,\;\e^2=P_{\s=\pi}(B_\perp)\e^1,  
\ee
This implies 
\be
Y^{mn}\G_{mn}\e^1=0,\;\;Y^{mn}\G_{mn}\e^2=0
\label{Ycond}
\ee
In the limit $\a'\rightarrow 0$ discussed in \cite{sw}, the leading term of
(\ref{Y}) is,
\be
Y^{mn}\G_{mn} \sim \left(\fr{1}{B}\right)^{mn} \G_{mn}
\ee
It follows that we have the following two conditions:
\bea
& \e_2 = \im \s^2\,\G_{4,...,9}\,\e_1 \nn\\
&  \left(\fr{1}{B} \right)^{mn}\G_{mn}\e_1=0
\label{stringresult}
\eea
The first equation breaks one half of the supersymmetry leaving sixteen
supercharges. The second equation breaks further leaving no supersymmetry in general.
We can apply the above discussion e.g., to the case of D5-D9 by replacing 
$\G_{4,...,9}$ with $\G_{6,...,9}$. The corresponding open string theory 
is supersymmetric only for $B$'s having special properties. 
For the D5-D9 case, the supersymmetry preserving $B$ is selfdual or 
antiselfdual and in this case the system preserves eight supercharges.
 
We also require that the boundary conditions for the bosonic
fields $X$'s are compatible with the supersymmetry transformations and obtain 
further constraints for the fermions $\theta$'s. It is easy to check
that the boundary conditions for fermions at both ends are 
\bea
& (\pa_\s + \pa_\t)\th_1=0, 
& (\pa_\s - \pa_\t)\th_2=0.
\eea
These conditions are consistent with light cone gauge and they solve
the last conditions in the equations
($\ref{sigma0}$) and ($\ref{sigmapi}$). 

\section{Gauge Theory Analysis}
In this section, we consider an  ${\cal N}=1$ abelian gauge theory in ten 
dimensions among which six directions are noncommutative.  
We use then the methods developed in \cite{gms,b,agms,hklm,hkl,gn} for 
noncommutative spaces to construct solitons in the six dimensional 
noncommutative gauge theory. We compare the supersymmetry breaking
pattern for these solutions 
with the one obtained in the previous section. We observe that 
they are equivalent to each other. 

The action for an ${\cal N}=1$ abelian gauge theory\footnote{For 
a ordinary abelian gauge theory, it is a partial 
derivative, $ \pa_{M} $ that acts on $\Psi$. For a theory on noncommutative
spaces the fields do not commute and it is necessary to replace a 
partial derivative by a covariant derivative.}
in $D=10$ is
\be
-S= \int d^{10}x \sqrt{G} \left( \fr{1}{4} \;F_{MN}F^{MN}
+\fr{1}{2}\bar{\Psi}\G^M D_M \Psi \right)
\ee
Since we take the same B-field configuration as in the previous
section, six dimensions become non-commutative. The
non-commutativity is parameterized by
\be
 [x^4,x^5] \equiv i\Theta^{45} \equiv i\th^{1}, \quad
[x^6,x^7] \equiv i\Theta^{67} \equiv i\th^{2}, 
\quad [x^8,x^9] \equiv i\Theta^{89} \equiv i\th^{3} 
\ee
It is  also convenient to introduce a complex coordinate system defined by
\be
 z^1=\fr{x^4+\im x^5}{\sqrt{2}}, z^2=\fr{x^6+\im x^7}{\sqrt{2}},  
         z^3=\fr{x^8+\im x^9}{\sqrt{2}} 
\ee
and define creation and annihilation operators,
\be
\hspace{1.5in} 
a^\dagger_p \equiv i\Theta^{-1}_{p\barq} \bar{z}^{\barq}\;, \; 
a_{\barp} \equiv -i\Theta^{-1}_{\barp q} {z}^{q},  
 \hspace{.5in} p=1,2,3 
\ee
Here we have chosen the annihilation and 
creation operators in accordance with the positivity
of the $\theta_{1,2,3}$. Namely for positive $\theta$'s the above
notation is valid, whereas for negative $\theta$'s we have to switch
the operators. We assume from now on that our $\theta$'s are all 
positive. In the ten dimensional action, we replace the integration 
over the six noncommutative dimensions with a trace over the Hilbert 
space acted by the creation, annihilation operators. We also replace the 
covariant derivative and the gauge field strengths on the noncommutative 
space in an appropriate way:
\bea
& D_p \rightarrow -[C_p,\ ], \nn\\
& F_{p\bar{q}} \rightarrow \im [C_p,\bar{C}_{\bar{q}}]-
\Th^{-1}_{p\bar{q}}, \nn\\
& F_{p q} \rightarrow -\im [C_p, C_q], \nn\\
& F_{\bar{p} \bar{q}} \rightarrow -\im [\bar{C}_{\bar{p}},
\bar{C}_{\bar{q}}]
\eea
where $C_p = -\im A_p+a^\dagger_p$. We consider, for simplicity,
that the metric on the noncommutative directions takes the simple
form: $G_{p \bar{q}}=\delta_{p \bar{q}}$. We make the above 
replacements in the ten dimensional action and obtain:  
\bea
S &=&-\fr{(2\pi)^3 \sqrt{-\mbox{det}\Th}}{g^2_{YM}} \int \sqrt{G}d^4x  
\mbox{Tr} \left( \fr{1}{4} F_{\m\n}F^{\m \n} 
+\fr{1}{2 }\left([C_p, \bar{C}_{\bar{q}}]
          +\im\Th^{-1}_{p\bar{q}}\right)
           \left([C_q, \bar{C}_{\bar{p}}]
          +\im\Th^{-1}_{q\bar{p}} \right)-\right.\nn\\
&&\left. - \fr{1}{2 } [C_p, C_q] [\bar{C}_{\bar{p}}, \bar{C}_{\bar{q}}]
+ D_\m C_p D^\m \bar{C}_{\bar{p}} 
+\fr{1}{2}\bar{\psi}\G^\m D_\m \psi + 
\fr{1}{2 }\bar{\psi}\G_p [\bar{C}_{\bar{p}},\psi]
 - \fr{1}{2 }\bar{\psi}\G_{\bar{p}} [C_p,\psi]  \right) 
\label{6daction}
\eea 
The supersymmetry transformations in six dimensions can be obtained 
from those of the ten dimensional theory by the same replacements we 
used for the action:
\bea
\d    A_\m  &=& \;\;\;\fr{1}{2}\bar{\eta}\G_\m \psi \nn\\
\d C_p &=& \;\;\;\fr{1}{2}\bar{\eta}\G_p \psi \nn\\ 
\d \psi    &=& -\fr{1}{4} \left( F_{\m \n}\G^{\m \n}+2D_\m C_p
                     \G^{\m p}+ F_{m n} \G^{m n} \right) \eta 
\label{6dsusy}
\eea
where $m,n=p,\bar{p}$ in the last equation. This transformations
have the same form even when we leave the metric $G$ arbitrary.

We consider now the solitonic solutions presented in \cite{agms} and 
establish when this solutions preserve supersymmetry. The solutions
have nontrivial values for $C_p$ and do not depend on the commutative 
coordinates:
\bea
& C_p=S^\dagger a^\dagger_p S, \; \bar{C}_{\bar{p}}=S^\dagger a_p S,
\label{sol}
\eea
The operator $S$ has the property that $S S^\dagger=1$ and 
$S^\dagger S=1-P_\circ$, where $P_\circ$ is the projector on the vacuum
$|0>$ ($a_{p}|0>=0$ for any $p$). As such, it is a 
static solution and has a single unit of topological charge. The solution
has field strength in the noncommutative directions: 
$F_{p\bar{q}}=-P_\circ \Theta^{-1}_{p\bar{q}}$. 
Solutions with arbitrary positive charge $m$ are obtained by simply
replacing $S$ with $S^m$. The discussion for the single 
charged solutions applies equally well for $m>1$. The first two
equations of (\ref{6dsusy}) are trivially satisfied. The first and
second terms of the $\psi$-variation vanishes once the 
solution is substituted. Therefore the only non-trivial part is the
last term,
\be
\d \psi =-\fr{1}{4}F_{p\barq}\G^{p\barq}\epsilon=0 
\label{gaugesusy}
\ee
 Since the indices in (\ref{stringresult}) are contracted by using
the closed string metric, $g_{mn}$, we need to convert
(\ref{gaugesusy}) into an expression where the indices are also
contracted with respect to the close string metric. Using the
solution, one can rewrite (\ref{gaugesusy}) as
\be
\d \psi =    \fr{1}{4} P_0 {\Th^{-1}}_{p\barq}
            {E^p}_{c} {E^{\barq}}_{\;\;\bard}\; {\G}^{c\bard} \epsilon 
\label{gaugesusy2}
\ee
where $c$ and $\bard$ ($p$ and $\barq$) are flat (curved) indices. 
$E^p_{\;\;c}$ represents the transverse components of the inverse 10-beins 
of the open string metric. They are related to the corresponding 
components of the closed string 10-beins, $e_p^{\;\;c} $, by \cite{sw}   
\be
E_p^c=\im (2\pi\a')B_{p\bar{s}}e^{\bar{s}c}
\label{6beins}
\ee
Substitution of (\ref{6beins}) into (\ref{gaugesusy2}) leads
to
\be
\d \psi=\fr{1}{4\pi^2{\a'}^2}P_\circ\left(\fr{1}{B}\right)^{mn}
                     \G_{mn}\,\epsilon =0 
\label{gaugeresult}
\ee
where now the contractions are with respect to the closed string
metric. Therefore this condition is equivalent to the one obtained 
in the previous section.

The result shows that in gauge theory we can construct supersymmetric 
solutions by the above method only in the case where the 
noncommutative parameters satisfy some particular relations. For
generic $\Th$'s the solutions do not preserve supersymetry. 
In the case of four noncommutative dimensions, by a similar analysis, 
we conclude that such instantons preserve supersymmetry only for 
selfdual or antiselfdual noncommutative parameter. 
\section{Fluctuations}

For properly chosen values of $\Th$'s, one has residual supersymmetry.
By study fluctuations around the solitonic 
configuration, we show here that tachyon does not appear. From
the action $(\ref{6daction})$ we can write the potential for the 
bosonic fields $C$'s:
\be
V(C) = \fr{1}{2}\mbox{Tr}\left(([C_p, C_{\bar{q}}]+i\Th^{-1}_{p\bar{q}})
   ([C_q, C_{\bar{p}}]+i\Th^{-1}_{q\bar{p}}) -
   [C_p, C_q][C_{\bar{p}}, C_{\bar{q}}]
\right)
\ee
where we use the summation over repeated indices $p$, $\bar{p}$. 
The above potential can be used also for the other cases of interest:
D0-D2 $\cite{agms}$ and D0-D4 if we consider $p,q=1$ and $p,q=1,2$ 
respectively. We consider fluctuations $C_p=C^0_p+\delta C_p$, around 
the solitonic solution: $C^0_p=S^\dagger a^\dagger_p S$ and decompose 
them in terms of the projection operators $P_\circ$ and $1-P_\circ$:
\bea
& \delta C_p=A_p+\bar{T}_p+W_p+ S^\dagger D_p S, \nn\\
& A_p=P_\circ \d C_p P_\circ,\; \bar{T}_p= (1 - P_\circ) \d C_p P_\circ,\\    
& W_p= P_\circ \d C_p (1-P_\circ), \; S^\dagger D_p S = (1-P_\circ)\d C_p
(1-P_\circ),\;\\ 
& S=|(0,..,0)><(1,0,..0)|+...,\nn
\eea
There is a corresponding decomposition for the hermitian conjugates of 
these. Focusing only 
on the part of this potential that can give a tachyon, we need 
only the first term in the expression of S. The first 
order term in fluctuations is zero and the second order term splits
in two independent terms, one for $T$'s and $W$'s and the other for 
$D$'s. We keep the one for $T$'s and $W$'s because the tachyonic
mode will come from this:
\bea
& V_2(T,W)=\mbox{Tr}(2\,\im\,\Th^{-1}_{q \bar{p}}(W_p \bar{W}_{\bar{q}}-
T_{\bar{q}}
\bar{T}_p)+(C^\circ_p\bar{C}^\circ_{\bar{p}}+\bar{C}^\circ_{\bar{p}}C^\circ_p)
(\bar{W}_{\bar{q}}W_q + \bar{T}_q T_{\bar{q}})+\nn\\
& -\bar{C}^\circ_{\bar{p}} \bar{C}^\circ_{\bar{q}}\bar{T}_p W_q -
C^\circ_p C^\circ_q \bar{W}_{\bar{p}} T_{\bar{q}}-
\bar{C}^\circ_{\bar{p}} C^\circ_q \bar{T}_p T_{\bar{q}}-
C^\circ_p \bar{C}^\circ_{\bar{q}}\bar{W}_{\bar{p}} W_q),
\eea
Using the simplified notations $\theta_{1,2,3}$, we obtain the
following potential in second order:
\bea
& V_2(T,W)=\mbox{Tr}\left(
- (W_p S^\dagger a_{\bar{p}} + T_{\bar{p}} S^\dagger a_p^\dagger)
( a^\dagger_q S \bar{W}_{\bar{q}}  + a_{\bar{q}} S \bar{T}_q) +
\right. \nn\\
& \left. + T_{\bar{p}} S^{\dagger}(a^\dagger_k\,a_{\bar{k}} +
a_{\bar{k}}\,a^\dagger_k) S \bar{T}_p + W_p\,S^\dagger
(a^\dagger_k\,a_{\bar{k}} + a_{\bar{k}}\,a^\dagger_k) S
\bar{W}_{\bar{p}} + \fr{2}{\theta_k} ( W_k\,\bar{W}_{\bar{k}} -
T_{\bar{k}}\,\bar{T}_k)\right) 
\label{pot2}
\eea  
We introduce in the expression for the potential the matrix 
components for the operators $T$'s and $W$'s :
\bea
& T_p = T^1_p|(0,0..)><(1,0..)| + \cdots \nn\\
& \bar{W}_p = \bar{W}^1_p|(1,0..)><(0,0..)| + \cdots
\eea
where by $...$ we mean other states of type $|0><|$ or $|><0|$ 
respectively. As can be seen from (\ref{pot2}), all the modes 
except $T^1_p$ will have positive mass. The mass terms for $T^1_p$ modes are:
\be
\sum_p |T^1_p|^2 \left(\sum_k \fr{1}{\theta_k}-\fr{2}{\theta_p}\right).  
\ee
We observe that for the D0-D2 system studied in \cite{agms},
($p=1$) we always have a tachyon. The condition for not having 
a tachyon in the case of D0-D4 is that 
$\theta_1=\theta_2$, namely the selfduality condition 
for $\Th$. For our case, the conditions are:
$\sum^3_k\fr{1}{\theta_k}-\fr{2}{\theta_p}\ge 0$ for any
$p$. The supersymmetry condition for our solution implies
indeed that there is no tachyon in the system.

\section{Conclusion}
 
The best studied example of AdS/CFT correspondence \cite{mal,gkp,w1},
the duality
between IIB supergravity and ${\cal N}=4, d=4$ SYM theory, was
motivated in \cite{iyp} by taking the viewpoint that there are two 
different but dual 
string theory descriptions of the same objects, D3-branes. The
viewpoint seems to point toward the duality between 
the two stringy descriptions themselves, i.e., between  IIB closed superstring 
description and oriented open superstring description. 

It was suggested in the same paper that in each description open
strings, closed strings and D3 branes may not all appear `explicitly'. For 
example, it was speculated that in the IIB closed string description 
the presence of open strings on the D-branes might be only associated 
with the size of
non-extremality of a D-brane soliton solution. Similar subtleties may lie
in the open string description for the realization of closed
strings. A related discussion can be found in \cite{sen,ghy} for open
string field theory in tachyonic vacua. 

As far as D-branes are concerned, open string theory has an efficient
way of realizing them: they appear as Dirichlet boundary 
conditions\footnote{It
was argued in \cite{iyp} with evidence \cite{dt,pst} that in the
context of AdS/CFT D-branes may
not merely be boundary conditions but provide a curved
background for open strings to propagate in. However, we did not, in
this article,
compare the open string description of D-branes with the realization
of them as a supergravity solution, but have remained
within the open string/gauge theory description. Therefore, the
background was taken to be flat for simplicity.}. As is well known,
Dirichlet boundary conditions are obtained through T-duality. 
Since the duality connects two equivalent descriptions, it is natural
to ask whether open string theory has another way of realizing D-branes.

With this motivation, we have considered the open string description of 
D3-D9 system in the Green-Schwarz (GS)
formulation with the non-zero components of the B-field transverse to
the D3 brane. The boundary conditions of the D3-D9 system gives an extra
condition for residual supersymmetry. 
Then we considered the supersymmetric gauge theory in ten dimensions and 
its dimensional reduction to six spatial dimensions. Turning on a constant 
B-field in the six dimensions, the gauge theory becomes
non-commutative. We have shown that the condition for residual supersymmetry 
are equivalent, in an appropriate limit, to that of the 
D3-D9 system mentioned above. This 
can be viewed as a generalization of the previous results in the
literature concerning lower codimensional cases where various
(non-commutative) solitons were identified as D-branes. \\

\noindent{\bf Note Added}: Sometime after our paper was published, a paper \cite{w3}
appeared that has some overlap with our results.

\end{document}